\documentclass[aps,pra,abstract, column]{revtex4}

\usepackage{amsmath}

\begin{document}

\title{Nature, Science, Bayes' Theorem, and the Whole of Reality}

\author{Moorad Alexanian}
%\email[]{alexanian@uncw.edu}

\affiliation{Department of Physics and Physical Oceanography\\
University of North Carolina Wilmington\\ Wilmington, NC
28403-5606\\}

%\date{\today}

\begin{abstract}
\[
\textbf{Abstract}
\]
A fundamental problem in science is how to make logical inferences from scientific data.  Mere data does not suffice since additional information is necessary to select a domain of models or hypotheses and thus determine the likelihood of each model or hypothesis. Thomas Bayes' Theorem relates the data and prior information to posterior probabilities associated with differing models or hypotheses and thus is useful in identifying the roles played by the known data and the assumed prior information when making inferences. Scientists, philosophers, and theologians accumulate knowledge when analyzing different aspects of reality and search for particular hypotheses or models to fit their respective subject matters. Of course, a main goal is then to integrate all kinds of knowledge into an all-encompassing worldview that would describe the whole of reality.

A generous description of the whole of reality would span, in the order of complexity, from the purely physical to the supernatural. These two extreme aspects of reality are bridged by a nonphysical realm, which would include elements of life, man, consciousness, rationality, mental and mathematical abstractions, etc. An urgent problem in the theory of knowledge is what science is and what it is not. Albert Einstein's notion of science in terms of sense perception is refined by defining operationally the data that makes up the subject matter of science.  It is shown, for instance, that theological considerations included in the prior information assumed by Isaac Newton is irrelevant in relating the data logically to the model or hypothesis.  In addition, the concepts of naturalism, intelligent design, and evolutionary theory are critically analyzed.  Finally, Eugene P. Wigner's suggestions concerning the nature of human consciousness, life, and the success of mathematics in the natural sciences is considered in the context of the creative power endowed in humans by God.
\end{abstract}
%\pacs{ 42.50.Pq, 42.50.Ex, 42.50.Dv, 03.67.Mn}
\maketitle {}

\section{Introduction}
In 1950, Albert Einstein gave a remarkable lecture (Einstein 2005, 46-8) to the International Congress of Surgeons in Cleveland, Ohio.  Einstein argued that the 19th-century physicists' simplistic view of Nature gave biologists the confidence to treat life as a purely physical phenomenon.  This mechanistic picture of Nature was based on the casual laws of Newtonian mechanics and the Faraday-Maxwell theory of electromagnetism.  These causal laws proved to be wanting, especially in atomistic phenomena, which brought about the advent of quantum mechanics in the 20th-century.

Einstein indicates that there are three principal features which science has firmly adhered to since Galileo Galilei.  This includes not only physics, but also all natural sciences dealing with both organic and inorganic processes,\\

``First: Thinking, alone, can never lead to any knowledge of external objects. Sense perception is the beginning of all research, and the truth of theoretical thought is given exclusively by its relation to the sum total of those experiences.\\

Second: All elementary concepts are reducible to space-time concepts. Only such concepts occur in the `laws of nature.' In this sense, all scientific thought is `geometric.' A law of nature is expected to hold true without exceptions; it is given up as soon as one is convinced that one of its conclusions is incompatible with a single experimental fact.\\

Third: The spatiotemporal laws are complete. This means, there is not a single law of nature that, in principle, could not be reduced to a law within the domain of space-time concepts. This principle implies, for instance, the conviction that psychic entities and relations can be reduced, in the last analysis, to processes of a physical and chemical nature within the nervous system. According to this principle, there are no nonphysical elements in the causal system of the processes of nature. In this sense, there is no room for `free will' within the framework of scientific thought, nor for an escape into `vitalism.' " (Einstein 2005, 46-8).\\

Einstein understands by science the study of the physical aspect of Nature and denies the metaphysical doctrine of vitalism, viz. that living organisms possess a nonphysical aspect that gives them the property of life.  In addition, Einstein invokes a reductionist view that the laws of physics and chemistry alone suffice to explain life functions and processes.  Of course, one cannot discern from Einstein's speech if he ascribes to the thesis that life emerges from a complex combination of organic matter, or more drastic still, inorganic matter.

It is clear that the inferences that Einstein makes follow from a combination of known data and assumed prior information.  Actually, the distinction one makes between what constitutes data and what constitutes prior information, or priors for short, is purely logical.  Any additional information needed to make an inference beyond that provided by the data is by definition prior information.  These concepts will be the topic of the next section.

\section{Mathematical Model of Reasoning}

The special theory of relativity is an often-cited example of inference based on purely deductive reasoning.  The constancy of the speed of light in vacuum plus the invariance of the laws of physics for different inertial frames suffice to derive, together with the assumed uniformity and isotropy of spacetime, all the results in the special theory of relativity.  Of course, inferences in science are mostly based on inductive or plausible reasoning where probability theory, not needed in deductive reasoning, is essential in the optimal processing of incomplete information as emphasized by Edwin Thompson Jaynes (Jaynes 2003, xix-xxix).  It was the problem of how the longitudes of Jupiter and Saturn had varied over long times that led Pierre-Simon Laplace to the study probability theory. Although the latter was based on consideration of games of chance, Laplace saw early on that it would become the most important object of human knowledge. The following  two quotes from Théorie Analytique des Probabilités, 1812 indicate Laplace's insight into the workings of the human mind, ``The most important questions of life are indeed, for the most part, really only problems of probability," and ``Probability theory is nothing but common sense reduced to calculation."
The problem of induction has a long history.  ``Philosophers have argued over the nature of induction for centuries.  Some, from David Hume (1711-76) in the mid-18th century to Karl Popper in the mid-20th century, have tried to deny the possibility of induction, although all scientific knowledge has been obtained by induction.  D. Stove (1982) calls them and their colleagues `the irrationalists' and tries to understand (1) how could such an absurd view ever have arisen?  And (2) by what linguistic practices do the irrationalists succeed in gaining an audience?"  (Jaynes 2003, 310).

Scientists do not test their theories against an infinite number of alternatives in order to establish the absolute status of a hypothesis embedded in the universe of all conceivable theories.  Instead, ``The functional use of induction in science is not to tell us what predictions must be true, but rather what predictions are most strongly indicated by our present hypotheses and our present information?"  (Jaynes 2003, 310).

Jaynes (2003, 24) develops quantitative rules of inference by working out the consequences of the following desiderata:\\

(I)    Representation of degrees of plausibility by real numbers.
\newline

(II)   Qualitative correspondence with common sense.
\newline

(III)   Consistency.
\newline

The fundamental principle underlying all probabilistic inferences is ``To form a judgment about the likely truth or falsity of any proposition $A$, the correct procedure is to calculate the probability that $A$ is true $P(A|E1E2 \cdots)$ conditional on all the evidence $E1 E2 \cdots$ at hand." (Jaynes 2003, 86).  The symbol $AB$, the logical product or the conjunction, denotes the proposition that both $A$ and $B$ are true. The arguments $A$ and $B$ in the formal probability symbol $P(A|B)$, which is a real number, are propositions and not numerical values.  One deals always with a finite set of propositions and extension to infinite sets is permitted only when this is the result of a well-defined and well-behaved limiting process from a finite set.

\subsection{Bayes' Theorem}

As indicated by Laplace, what one wants is to use probability theory to indicate which of a given set of hypotheses $\{H1, H2,\cdots\}$ is most likely to be true in the light of the data and any other evidence at hand.  For example, the hypotheses may be various suppositions about the physical mechanism that is generating the data.  Fundamentally, physical causation is not an essential ingredient of the problem; ``what is essential is only that there be some kind of logical connection between the hypotheses and the data."  (Jaynes 2003, 88-9).

If one uses probability theory as logic, then Bayes' theorem, first found by the amateur mathematician the Rev. Thomas Bayes (1763), can be used to make probabilistic inferences from incomplete information, say, data from the observational or the experimental sciences, viz.
\[
P(H|DX) =P(H|X) P(D|HX) /P(D|X),
\]
where $P(A|E)$ stands for probability that $A$ is true conditional on the evidence $E$ and\\
\newline
$X$= prior information,\\
\newline
$H$ = some hypothesis to be tested,\\
\newline
$D$= data.
\newline

If the physical evidence $D$ considered by different people is the same, then the real question is, what prior information $X$ do different people consider in making logical inferences?  The posterior probability $P(H|DX)$ depends on the product $DX$ and not separately on the data $D$ and the prior information $X$.  This indicates that the separation of the evidence into data and prior is one of convenience when organizing a chain of inferences.  In the study of the physical universe, one may suppose that the prior information $X$ be expressible in purely mathematical and physical terms.  Of course, human consciousness and rationality enters in how and what type of data $D$ is obtained and what sort of prior information $X$ is used.  However, a robot that reasons according to certain definite rules designed by us can make the reasoning and deduce the consequences that follow from Bayes' theorem, once such choices are made.

Consider the following application of Bayes' theorem (Kahneman and Tversky 1973). Prior information:  $X$ = ``In a certain city, 85\% of the taxicabs are blue, 15\% are green ." Data: $D$ = ``A witness to a crash who is 80\% reliable (i.e., who in the lighting conditions prevailing can distinguish correctly between green and blue 80\% of the time) reports that the cab involved was green." In this psychological test, the subjects are then asked to judge the probability that the cab was actually blue. The correct result for the probability that the cab was actually blue $P(B|DX)$ is given by Bayes' theorem, where the hypothesis H is either B or G. Now the prior information gives that $P(B|X) = 0.85$ and $P(G|X) = 0.15$. In addition, the data D, given the hypothesis H and prior information $X$, yield $P(D|BX) = 0.20$ and $P(D|GX) = 0.80$.  The probability $P(D|X)$, needed in Bayes' theorem, follows from the normalization condition $P(B|DX) + P(G|DX) =1$. Thus, $P(D|X) = P(B|X) P(D|BX) + P(G|X) P(D|GX) = (0.85)( 0.20) + (0.15)(0.80) = 0.29$. Therefore, by Bayes' theorem, $P(B|DX) = P(B|X) P(D|BX)/P(D|X) = (0.85)(0.20)/(0.29) = 0.59$.

It is interesting that the subjects in the psychological test tended to guess a value for $P(B|DX)$ of about 0.20, which corresponds to ignoring the prior information $X$, viz., $P(B|X) = P(G|X) = 1/2$,  thus $P(D|X) =1/2$ and $P(B|DX) = P(D|BX) = 0.20$ as suggested by the data alone. An equally yet opposite error would be to cling solely to the prior  information at the expense of the data, which may be common in certain circumstances in religion and politics or, at times, even  in science. In this instance,  $P(D|GX) = P(D|BX) = 1/2$ and so $P(B|DX) = P(B|X) = 0.85$. These two extreme cases make either the prior information or the data $D$ irrelevant, thus drastically changing the inferences made.
Human reasoning, if unfettered, can bring into a problem of inference prior information that is irrelevant to the data being considered.  ``Obviously, the operation of real human brains is so complicated that we can make no pretense of explaining its mysteries; and in any event we are not trying to explain, much less reproduce, all the aberrations and inconsistencies of human brains.  That is an interesting and important subject; but it is not the subject we are studying here.  Our topic is the normative principles of logic, and not the principles of psychology or neurophysiology."  (Jaynes 2003, 8).

The introduction of an imaginary being, a robot, is an attempt to guide human reasoning about propositions according to the given desiderata.  The making of inferences must be based on a clear connection between the data one has plus the relevant prior information one uses in making logical inferences.  Otherwise, data or prior information not explicitly stated may be used unknowingly. The best illustration of this caveat is provided by David Hilbert's systematic study of the axioms of Euclidean geometry, which led Hilbert to propose 21 such axioms rather than the usual 5 axioms of Euclid of Alexandria.   Hilbert published Grundlagen der Geometrie in 1899 putting geometry in a formal axiomatic setting.

It is important to remark that assigning prior probabilities to verbal statements, e.g., concerning purpose, design, intelligence, consciousness, etc., is an open-ended problem of logical analysis since there is no unique way to accomplish that.  In fact, prior probabilities represent a state of knowledge rather than a physical fact.  Presently, there are four general principles (Jaynes 2003, 88-9) for assigning priors, viz., group invariance, maximum entropy, marginalization, and coding theory.  For instance, the question of macroscopic final causes, e.g., arranging a physical system in order to attain a given temperature and pressure, are accomplished by using the principle of maximum entropy.  Of course, physicists have succeeded in handling also physical systems in order to obtain particular microscopic final causes.  It is in this sense that there is indeed purpose and design in the Universe.  However, the latter is a theologically neutral observation that does not invariably lead to the existence of supernatural contravention of the laws of physics.  (Jaynes 2003, 614).

\subsection{Hypothesis Testing}

The fundamental question in science, and for that matter, rational thinking in general, is how to decide between differing available models or hypotheses in order to account for the existing data.  Of course, one has to be rather clear on what data is being considered and how the data is collected in that particular kind of knowledge.  One best way to avoid confusion is to characterize a kind of knowledge according to a subject matter.  The question is how does one do that in an unambiguous, operational fashion so that it is clear what kind of evidence would be necessary in order to establish the truth or falsehood of claims being made.  Of course, what prior information is used to analyze such data is crucial.

In order to keep track of what assumptions are being made in the process of developing scientific theories of Nature, one must specify what the data $D$ is that one is taking into account and what prior information $X$ one is assuming.  Of course, the division of what is data and what is prior information is a purely logical one.  In any case, any additional information beyond that provided by the data of the current problem is by definition prior information.  Bayesian methods are used whereby the mathematical rules of probability provide consistent rules for conducting inferences.  Einstein indicates that physical influences can propagate only forward in time (causality).  However, logical connections, which may or may not correspond to causal physical influences, propagate equally well in either direction, viz., ``one man's prior probability is another man's posterior probability."  (Jaynes 2003, 88-9).

In Bayes' theorem, the logical product, say, $HX$ denotes the proposition that both propositions $H$ and$X$ are true.  Therefore, in order to know the likely truth of $H$, viz. $P(H|DX)$, given the data $D$ and prior $X$, one needs not only the sampling probability $P(D|HX)$ but also the prior probabilities $P(D|X)$ for$D$ and $P(H|X)$ for $H$.  All probabilities are necessarily conditional on the prior information $X$.  If parts of$X$ are irrelevant to the problem at hand, then such parts of $X$ will cancel out mathematically and so $P(H|DX)$ will not depend on such parts of $X$ and so are truly irrelevant in determining the likely truth of $H$.

\section{The Science of Isaac Newton}

Humans are puzzled by two fundamental thoughts: the question of origins and/or existence and what sustains and/or controls that which already exists.  Some may further question if entities can come into being that are not causally related to what already exists.  Biblical answers to these questions are, the creation account by God (Genesis 1, 2) via Christ (John 1:3), Christ sustains the creation by the word of His power (Hebrews 1:3), and God entering His creation through the Incarnation of His Son (John 1:14).  However, such answers still do not quench human curiosity and so further progress is attempted by the methods of science that may still be guided by theological considerations.

Blaise Pascal summarized the human dilemma succinctly, ``For in fact what is man in nature? A Nothing in comparison with the Infinite, an All in comparison with the Nothing, a mean between nothing and everything. Since he is infinitely removed from comprehending the extremes, the end of things and their beginning are hopelessly hidden from him in an impenetrable secret; he is equally incapable of seeing the Nothing from which he was made, and the Infinite in which he is swallowed up." (Pascal 1941, 23).
The development by Isaac Newton of his gravitational theory illustrates how suppositions enter in our attempts to study Nature.  In constructing his theoretical model of the solar system, Newton said nothing about the origin of the solar system; he simply assumed its existence and was concerned only with planetary motion.  Is the origin of planets outside the confines of science? Of course not!  However, if one wants to study the origin of planets, then one must consider a more encompassing model with new initial conditions that will allow predictions regarding the existence of planets and, in addition, successfully describe their motion as Newton accomplished.

The process of extending the usefulness and predictability of a model can go on and on encompassing more and more of the description of Nature and leading to more predictions than given by previous theories.  However, eventually one reaches a point where no further progress is possible and one would achieve the elusive, so-called Theory of Everything, which purportedly explains and links together all known physical phenomena.  The suppositions and initial conditions made in this ultimate theory will invariably lie outside science and into the realm of the purely metaphysical.  John Wheeler (Wheeler and Ford 1998, 262) asked, ``How come existence?" and its subsidiary, ``How come the quantum?" as an attempt to incorporate some deeper concepts into science.  Wheeler's ``It from bits" is an attempt to answer some of these fundamental questions but still within the confines of a science based on materialism.

The data used by Newton in constructing his gravitational theory, obtained by the Danish astronomer Tycho Brahe, is summarized in Johannes Kepler's three laws that rule the motion of the planets around the Sun.  What prior information did Newton use?  Surely, Newton used all he knew and, in addition, created the calculus in order to derive the gravitational force between two spherically symmetric massive objects.  Was all the prior information Newton had at hand relevant for the conception and construction of his theory of gravitation?  Certainly not!  Much of the prior information Newton had was irrelevant to his discovery since one is not here dealing with the origin of ideas or the historical development of a theory.  Rather, one is seeking the logical connection between the model or hypothesis and the data.

Clearly, more than one model or hypothesis may be connected logically with the data.  For instance, the Ptolemaic system, with the aid of epicycles, explained the apparent motion of the Moon, Sun, and the retrograde motion of the planets.  In fact, more levels of epicycles (circles within circles) were added to the models to match more accurately the existing data.  The important question is how does one compare different conceivable models, distinguishes their individual merits, and selects the best?  The stakes are indeed high since a false premise built into a model that is never questioned cannot be removed by any amount of new data.  The comparison of models in order to choose the best amongst a prescribed class of alternative hypotheses can be based on the principle of reasoning through modern Bayesian analysis.   In fact, the Franciscan Monk William of Ockham may be considered a forerunner of modern Bayesians.

It is interesting that Newton reconciled or integrated his scientific work with his theological beliefs: ``A God who had been active in history, and who had communicated with mankind through the prophets, also might be expected to be active in Nature.  When Newton was in a more speculative frame of mind, he often seemed to be exploring strategies that would heighten that sense of divine dominion and control.  It would seem reasonable to suppose that this was one reason why he was reluctant to make the gravitational force an innate property of matter... Newton wrote, `There exits an infinite and omnipresent spirit in which matter is moved according to mathematical laws.' "  (Brooke 1988, 169-83).

According to Newton, the biblical God interacts also with the physical aspect of His Creation via purely physical agencies: ``Newton's universe needed to be tinkered with from time to time.  He thought that the gravitational attraction of the planets would cause their orbits to accumulate deviations, and the constant radiation of matter from the solar system would have to be made up.  (The degeneration of the cosmos is not merely the result of friction in the interplanetary æther.)  As much as possible he introduced natural agencies, such as the passage of comets through the solar system, as the means by which God made corrections and restored mass."  (Wilder 1991, 2-16).

If the central problem for Newton was a theory that led to Kepler's three laws, then clearly invoking God amongst his prior information was certainly not relevant in order to make the logical connection of the model or hypothesis to the data.  It is important to distinguish between the scientific model proposed and the actual, real solar system.  A scientific model need not refer or invoke a Creator; however, all that is may be the action of a Creator, which both creates and sustains all that exists.

Despite Newton's attempt of integrating his theology with science, his principle of universal gravitation is clearly a physical theory designed to describe the physical aspects of the solar system. Of course, any theory of the formational history of the universe would involve the whole of reality, which would include not only the purely physical, but also the existence of conscious, rational human beings.  The fundamental question then becomes what kind of model or hypothesis, initial conditions, and prior information are suitable and sufficient for such an all-encompassing description of Nature?

\section {Science and the Whole of Reality}

It is difficult to carry on constructive dialogue without defining the meaning of concepts like ``reality," ``natural," ``science," etc.  What is ``natural" may mean all that exists in Nature but then what is ``Nature"?  One must have operational definitions of these concepts, as is the norm in the physical sciences, in order to attain objectivity.  Witness the notion of time and the operational definition of time via clocks, which are purely physical devices.  If such conceptual order cannot be established in physics, then confusion will reign in the study of all aspects of reality.  For instance, philosophers have argued for millennia on the nature of time, in particular, whether time is real or a mere appearance.  As Einstein would indicate, just thinking of what is time does not suffice and so time must be defined with the aid of physical clocks that establishes an unambiguous, objective meaning of time, which allows knowledge of external objects and their temporal behavior.

Einstein (2005, 46-8) considers Nature to encompass both the organic and the inorganic and knowledge of natural phenomena is based on observations.  In addition, the laws governing the workings of Nature are reducible to elementary, spatiotemporal, physical concepts.  More importantly, Einstein denies the notion of the nonphysical in the causal development of Nature.  However, how are observations to be conducted in order to collect the necessary data in the study of Nature?  What entities or devices register observations as data?

The subject matter of science is indeed the physical universe as posited by Einstein.  It is clear that unless one defines science in terms of a subject matter, then all sorts of intractable confusion will arise as to the true nature of unadulterated science.  Therefore, the data that is constitutive of the subject matter of science can be collected, in principle, solely by physical devices (Alexanian 2000, 80).  This accords with Einstein's notion of science and makes precise his requirement that ``sense perception is the beginning of all research….the sum total of those experiences" (Einstein 2005, 46-8).  However, is the whole of reality physical?  It is not clear whether Einstein denies the existence of nonphysical elements in the whole of reality or only ``in the causal system of the processes of nature."

The metaphysical question regarding the nature of the whole of reality goes beyond the bounds of science (Alexanian 2002, 287-88).  I for one believe that knowledge of the true nature of humans, in particular the ability to know self and reason, involves more than the physical.  Surely, human consciousness is not detectable by purely physical devices and only the nonphysical self in humans can ``detect" or know consciousness.  Of course, different levels of conscious experience are related to brain-states but self cannot be reduced to such physical states.  Therefore, human consciousness is outside the purview of science (Alexanian 2002, 65), which allows humans to ``detect" nonphysical entities such as self, information, mental constructs, mathematical abstractions, etc.  It seems unlikely that the notion of life can be defined in terms of purely physical concepts and so it may be that the true essence of what life is lies outside the confines of science as well.

In the process of developing scientific theories, human consciousness and reasoning summarizes the data into laws and creates the mathematical models with predictive power that describe Nature.  However, the human elements of consciousness and rationality are not an integral part of the laws and models themselves.  That is to say, human consciousness and rationality are not needed to connect logically the data to models or hypotheses and so are irrelevant as prior information.  Theoretical models of Nature and the predictions that follow from them are exactly like mathematical systems with axioms and theorems.  Einstein indicates (2003, 46-8) that all scientific thought is ``geometric" and the abstract concepts exist only in conscious minds. It would be almost miraculous if a human being would be one of the predictions derived from a scientific theory envisioned by human beings themselves.

Scientific studies need not invoke God, i.e. God does not form part of the prior information or data, provided one does not deal with ontological questions, in particular, regarding the origin of the assumed initial conditions of our spatiotemporal laws and the laws themselves.   Additional prior information or data may be used to not only connect logically the new hypotheses with the original data but also to encompass additional prior information or data from a different kind of knowledge, say theology.  The latter would represent an attempt to integrate science and theology and thus create a synthetic discipline that would explain scientific knowledge and theological issues like purpose, meaning, value, etc.

A sensible definition of the whole of reality is then the realm of all data collected by purely physical devices and conscious, rational human beings.  Consequently, the subject matter that encompasses the whole of reality goes beyond the subject matter of science, viz. the physical, and includes the nonphysical as well.  Thus, the physical aspect of Nature is a subset of the whole of reality, which does not include the abstract concepts, like mathematics, information, etc. that exist in the human mind.  One is dealing here only with science, which I have defined unambiguously and operationally, and not with other kinds of knowledge. For instance, the study of theology and/or history would require different, additional data and prior information.  Of course, in the order of knowledge, ``Metaphysics and theology constitute the two domains of the ontological context.  Since metaphysics delimits the possible, and since theology may assert something as actual which from the standpoint of certain metaphysics is impossible, therefore there can be a conflict" (Martin 1957, 328-29).  Therefore, the incompatibility between materialism and theology ``arose not from metaphysics in a regulative function, but from metaphysics in a reductive function."  (Martin 1957, 328-29).

Many scientists are interested in the relationship of science and religion. In particular, Howard J. Van Till, a physicist and astronomer who is a founding member of the International Society for Science and Religion, is devoted to topics regarding the relationship of science and religion. Walter R. Thorson,  a theoretical chemist, has deep interest in the philosophy of science and related issues in theology. Their proposed naturalism  (Thorson 2004, 26-36) and (Van Till 2004, 292-5) are theistic owing to the prior information of a Deity and a Creator, respectively, in their initial conditions of their spatiotemporal laws.

Clearly, the existing physical data does not lead to a specific metaphysics.  Hence, choosing materialism as the correct metaphysics is a form of nihilism since it eliminates legitimate kinds of knowledge that deal with the nonphysical, say, theology, consciousness, etc.  The study of the time development of the universe deals with both physical and historical sciences. (Alexanian 2007, 10, 12).  The former are experimental in nature, whereas the latter are akin to forensic science, where the results of experimental sciences are used to validate an assumed sequence of unique events.  Surely, interferences from the physical data can easily go beyond purely scientific descriptions into theological considerations.  This was certainly the case in the work of Newton.  Similarly in the naturalism of Thorson (Thorson 2004, 26-36) and Van Till (Van Till 2004, 292-5), as well as, in intelligence design (ID) (see O'Leary 2004, chaps. 13-6), where either explicitly or implicitly the notion of a Supreme or Superior being is introduced.  However, such prior information is irrelevant in the description of the physical aspect of the universe where ontological questions are not addressed but may be necessary when dealing with the whole of reality---both physical and nonphysical. (Alexanian 2007, 85-86).  These approaches are discussed in the light of Bayes' theorem in the following section.

If materialism is equated to physicalism, then only the physical is real.  Now materialism is obviously wrong since it denies the existence of nonphysical, abstract concepts needed and used in science.  The very notion of creating theories to explain the physical goes beyond the physical owing to the need of nonphysical prior information.  For instance, there is no material or physical existence for the transcendental number?  However, such a concept is indispensable in theoretical science.  In addition, materialism denies the very notion of Deity, which transcends the physical yet is ``detected" by the nonphysical in humans. (Alexanian, 2006, 254-255).

\section{Naturalism, Intelligent Design, and Evolution}

The power of Bayes' theorem lies in making clear, in the logical chain of inferences, what constitutes the input, viz. data and prior information, and what constitutes the output, viz. the likelihood of a hypothesis from that of a given set.  Still, it is an open and difficult problem of how to translate different verbal prior information into numerical prior probabilities. This makes one less able to deduce the likelihood of a hypothesis in the historical sciences from the data and assumed prior information than is the case in the physical sciences and thus more difficult to appraise the validity of different hypotheses conditional on the data and prior information.  Nonetheless, Bayesian analysis allows us to clarify and evaluate the logical relation of different hypotheses to the data and the assumed priors.

The following simple example illustrates the use of Bayes' theorem in cases of verbal prior information. Consider the statement ``Doubt is an integral element of faith." How does probability come into the analysis of this statement? Surely, faith without doubt denotes certainty, which is contrary to its nature and indicates already a probability of one. What is the prior information that can account for the correctness of the statement? It is a matter of choice on what or on whom one puts one's faith. Therefore, the choice in the prior information is by necessity accompanied with doubts vis-à-vis the other, non-selected faiths. However, it is difficult to quantify such arguments. If all choices of differing faiths are equal, then one has Laplace's ``Principle of Insufficient Reason."

Scientific theories are like mathematical systems, say Euclidean geometry, with suppositions (axioms) and predictions (theorems). Of course, the logic can go both ways and so one can make some theorems axioms, etc.  There is no designer in the theory except the human who created the mathematical scheme. If one wants to explain humans that created the mathematics and developed the scientific theories, then one has to unify science with some other kind of knowledge. It is in this sense that one infers a Creator from what exists.

Thorson (Thorson 2004, 26-36) indicates that intelligent design (ID) is not a legitimate scientific agenda for biology and thus classifies ID as natural theology rather than a naturalistic approach to science.  By ``naturalism," Thorson means, ``that in science we deliberately refrain from using explanatory paradigms or concepts that appeal either to divine agency itself or to any direct surrogate for such agency"  (Thorson 2004, 26-36) . Thorson suggests a possible way ID may be made compatible with a ``naturalistic" approach to biology.  In particular, ID should focus on Thorson's "functional logic" that endows biological systems with ``a logical organization toward achievement of certain tasks or functions" (Thorson 2004, 26-36).   Such functional logic, although compatible with mechanistic logic, most likely is not derivable from it.  Thorson proposes, in order to emulate the successes of physical science, deferring the question of origins, which may actually go beyond the bounds of science.

On the other hand, Van Till believes that ID is ``inherently incapable of developing a positive case for any particular non-naturalistic worldview" (Van Till 2004, 292-5).  By ``naturalism," Van Till means, ``a world view (whether theistic or atheistic) that posits the sufficiency of the system of natural causes to bring about the actualization of the full spectrum of physical structures and biotic forms and systems that have appeared in the formational history of the universe" (Van Till 2004, 292-5).  Van Till proposes that our universe is gifted by its Creator with a ``robust formational economy."  This would lack ``nothing needed to actualize, without compensatory non natural action, every type of structure and organism that has appeared in its formational history."  Van Till distinguishes ``materialism" that denies the existence of any immaterial Deity from the theistic ``naturalism" that he posits.  However, in contrast to ID, which Van Till considers closely associated with Divine intervention, Van Till's naturalism comes dangerously close to Deism. The ``naturalisms" of both Thorson and Van Till includes as prior information the notion of a Deity and a Creator, respectively.  Yet, the data both are trying to explain are not theological but physical and biological in nature, and refer to a Deity or Creator to set up the initial conditions that purport to explain the history of the universe.

Evolutionary theory comes in two main flavors according to the assumed initial conditions.  One form supposes that initially there is nothing but nonliving matter, say, the Big Bang creation of the universe, and life and complex living organisms develop as the universe evolves by spatiotemporal laws.  This is materialistic metaphysics, which violates the order of knowledge, where metaphysical propositions are regulative rather than constitutive of historical propositions (Martin 1957, 318).  Alternatively, the initial state is taken to consist of simple, living organisms and the essential problem is how complex, living organisms arise in time.  Theories with explicit mechanisms and predictive power that would achieve this sort of temporal behavior are rather difficult to construct since it would involve a thorough knowledge of experimental science and would have to possess the ability to describe the historical development of our universe.

Both the ``functional logic" of Thorson and the ``robust formational economy" of Van Till are a form of evolutionary theory, albeit theistic.  Thorson (Thorson 2004, 26-36) relegates the question of origins to a Deity and Van Till (Van Till 2004, 292-5) ascribes his ``robust formational economy" to a Creator.  The theory of ID holds that certain features of the universe and of living organisms are best explained by an intelligent cause.  On the other hand, evolutionary theory is based on undirected processes such as natural selection and so it supposes no known teleological content either in the initial state or in the temporal development of the universe.  Therefore, it is not clear whether the intelligent cause in ID is solely in the initial state, as in Thorson and Van Till's theories, or else, as Van Till emphasizes (Van Till 2004, 292-5), it also appears in the dynamics of the system in the form of divine intervention.

Intelligent design is based on the inference that certain biological information must be the product of an intelligent cause.  This can be realized, for instance, if the designer arranges the information content of the initial condition so that the time development of the system would give rise to the observed complexity.  A stronger version of the latter would posit that no physical theory could give rise to irreducible complexity if not designed to do so.  If successful, then design would not rely on inference but would be a necessary condition of complex biological systems.  This endeavor would be the biological equivalent of G\"{o}del's incompleteness theorem in any axiomatic mathematical system (Nagel and Newman 1958).   In the case of ID, the domain would encompass all conceivable physical theories of Nature.  Accordingly, no hypothesis that does not include a designer in the prior information is able to explain the whole of biological data. The proof of such a proposition would constitute indeed a Herculean task.

Darwinism presupposes a purely materialistic description of Nature whereas ID supposes the existence of an intelligent designer.  As indicated above, ID cannot refute the naturalism of Thorson and Van Till.  In fact, the difference between ID and Thorson and Van Till's theories is that ID claims to identify by means of scientific data the originator of the initial conditions as an intelligent agent.

The whole gamut of possible initial states, which can range from the state of creation on the seventh day of Genesis 2 (or an earlier day) to the presently supposed Big Bang origin of the universe, are available to any theory.   It is important to indicate that Thorson's, Van Till's, and ID's efforts of explaining what presently exists are purported to be scientific theories.  Of course, in the true reductionist tradition of science, the goal is always to consider the simplest, least complex initial conditions together with a dynamical model that successfully describes the temporal behavior of an initially lifeless, physical universe.  Accordingly, in such scenarios, life, rationality, and consciousness would be a consequence of the temporal evolution of the universe.

\section{Science, Life, and Physics}

The division of Nature into the physical and nonphysical aspects raises the issue whether the nonphysical can be explained fully in terms of the physical or not. Humans possess both physical and nonphysical attributes and beyond Nature itself, there is the supernatural that transcends Nature.  Eugene Paul Wigner observed, ``There are phenomena what the present physics cannot describe.  One of them is the life. This circumstance is as peculiar as if physicists did not consider gravity.  However, gravity exists and life exists as well.  Humans are here with all of their thoughts, joys and desires.  It used to be said that humans obey the laws of physics and emotions do not count.  This saying reminds me of the fact that 100 years ago the physics textbooks stated: `Atoms may exist but that is irrelevant for physics.'  Some scientists say that emotions may influence the soul but not the motion of our hands or feet. However, it is hard to accept that we are only machines. Consciousness may influence the events in the same way as gravity does. This means, however, that in shaping the future of our universe something plays a relevant role what physicists are not interested about, in a way as they were not interested about atoms 100 years ago" (Wigner 1999). Wigner was a mathematical physicist and Nobel Laureate in Physics who laid the foundation for the theory of symmetries in quantum mechanics and commented, ``It was not possible to formulate the laws (of quantum theory) in a fully consistent way without reference to consciousness." If true, then the latter would provide a rather strong connection between the physical and the nonphysical aspects of Nature.

The notion whether life itself can be understood or explained in terms of purely physical concepts or not is a rather complex problem to which all fundamental problems in physics pale in comparison.  For instance, can life be the result of an emergent property just as wetness is an emergent property of water molecules when in the liquid state?  Thorson is skeptical that ``any purely physical model can explain biological organization" (Thorson 2004, 26-36).  Therefore, in essence, Thorson's concern with a naturalistic approach to biology represents extending the definition of what science is to include the concept of life, whatever life is. Surely, the notion of life can be used in science as an unknown, phenomenological concept its true nature awaiting final understanding much as Thorson suggests regarding questions of origin.

Wigner raises the question of human consciousness and life as possible scientific topics. Is it then possible to extend further the definition of science beyond the realm of the purely physical in order to encompass the notion of human consciousness as well? Alternatively, is human consciousness an emergent property of the purely physical?  Positive answers to these questions would imply that Thorson and Van Till's naturalisms and ID theory could be reduced to materialism, since there would be no need of a Deity to neither posit the initial conditions nor contravene the laws of physics.  However, this leads to an intrinsic incompatibility with theology since materialism ``does not allow the possibility of the truth of any strictly theological proposition" (Martin 1957, 326).  However, the undetectable nature of self by purely physical devices indicates that naturalism cannot encompass the whole of reality, even if one supposes the existence of life and/or humans in the initial conditions.  Consciousness may even be an emergent property of the physical yet the physical cannot detect it, much like the inability of a water molecule to detect the wetness of water, which would be rather puzzling.  This conclusion does not invoke the incompatibility of materialistic metaphysics and theology but relies on the existence of conscious beings that ``detect" nonphysical entities, even God, but are undetected by means of purely physical devices owing to the existence of nonphysical consciousness.

Wigner (1960, 1-14) also examined how mathematics accurately describes physical phenomena thus affirming Galileo's tenet that ``The book of Nature is written in the language mathematics."  Wigner asserts, however, ``.....the enormous usefulness of mathematics in the natural sciences is something bordering on the mysterious and there is no rational explanation for it."  It may be that mathematical descriptions of Nature work because the human mind creates mathematics.  Now God created both humans and Nature. The circle is thus completed with the creative power of humans mimicking that of God and understanding Nature by the power of their intellect.  Thus, the existence of self, which ``detects" the spiritual, exemplifies the image of God in humans and points to the existence of theological and mathematical truths innate to humans. This resembles somewhat the theory of occasionalism of Nicolas Malebranche who claimed that the only causal power is God (Malebranche 1997).

It is important to remark that the laws of experimental sciences, which underlie the laws of Nature, are generalizations of historical propositions and are quite consistent with most theological presuppositions. It is in the study of unique historical events, say in cosmology or abiogenesis, where the conflict, the so-called war between science and religion, truly arises. It is the historical presuppositions in the historical sciences, not the science used that is based on the findings of the experimental sciences, which gives rise to a possible discord.

For instance, the Christian faith is based on the historicity of God incarnate in Jesus of Nazareth and the redeeming power of the cross.  Experimental science has nothing to say regarding any particular historical event; least of all regarding the entering of the Creator Himself into His creation.

Question of origins, in particular, the origin man, pose a most difficulty problem. Note that even though the results of experimental sciences are used to analyze extant data, the problem of origins is essentially a historical, not a scientific problem. Additionally, it confronts Christians with how to reconcile the origin and present moral status of man vis-à-vis the redeeming grace found in the death and resurrection of Jesus the Christ.

``Then the Lord God formed man of dust from the ground, and breathed into his nostrils the breath of life; and man became a living being." (Genesis 2:7 NASB).  Physical science has successfully developed paradigms to study nonliving ``dust."  However, can science make the ``breath" of God part of its subject matter?  Is the concept of life so elusive that it becomes scientifically indefinable?  Perhaps the inability of nonliving matter to detect and identify life as well as consciousness indicates that only life itself can ``detect" and know life.  Similarly, only self can ``detect" and know self. Consciousness presupposes rationality, rationality presupposes life and life presupposes God.  Human rationality and consciousness is used to know both Nature and God yet paradoxically humans are unable to formulate a scientific theory of neither life nor self.

\section{Conclusions}

The purely physical is the subject matter of physics, the prototype of all sciences.  An extreme form of reductionism supposes that all that exists is the purely physical and the nonphysical aspect of reality follows from the purely physical and the laws governing their actions. This poses a gargantuan theoretical problem and so a more practical approach would built on less fundamental foundations by supposing the existence of more complex entities, even living beings, and explaining the remaining aspects of both the purely physical and nonphysical in terms of them.

Therefore, a definition of science in terms of a subject matter is necessary; otherwise, the term would be as equivocal and meaningless as the term evolution has become.  This is clearly not reductionism since one is not reducing all that exists to the physical; therefore, one is invoking neither physicalism nor materialism.  Many consider such a definition of science as excluding, for instance, ID as science.  However, if it does do that it would also exclude Darwinism as well.  A definition of science based on the study of the physical aspect of Nature is a two-edged sword that exposes the nonscientific, philosophical baggage of Darwinism.  If such a definition of science is unacceptable, then one has to extend the definition beyond that presently used by physicists, astronomers, chemists, neurophysiologists, molecular biologists, etc.  The question is how does one do that in an unambiguous, operational fashion so that it is clear what kind of evidence would be necessary in order to establish the truth or falsehood of claims being made in the name of science.

Surely, one can make higher order inferences from scientific data by using additional prior information than is required to do unadulterated science.  Such higher order inferences are normally made in order to integrate scientific data with evidence from other kinds of knowledge.  An obvious example of this is, ``The heavens are telling of the glory of God; And their expanse is declaring the work of His hands" (Psalm 19:1 NASB).  One needs scientific data and prior information that would encompass both scientific and theological considerations.

If living organisms and conscious beings are to be part of the subject matter of science, then how does one define these new concepts with some sort of scientific rigor?  Physicists have studied the notion of life, witness the works of Erwin Schr\"{o}dinger, Max Delbr\"{u}ck, etc., and are now considering the question of consciousness.  It is doubtful that the definition of science as the study of the physical universe will be able to encompass the difficult concepts of life and consciousness.  No one has any idea how the latter concepts can be made part of the scientific endeavor with the same rigor that characterized the concepts and science developed by Galileo, Newton, Planck, etc., whichever way science is defined.

\[
\textbf{REFERENCES}
\]
Alexanian, Moorad. 2000.  ``Teaching, Propaganda, and the Middle Ground." Physics Today 53, no. 11, 80.\\

Alexanian, Moorad. 2002. ``Physical and Nonphysical Aspects of Nature."  Perspectives on Science and Christian Faith 54, no. 4, 287-288.\\

Alexanian, Moorad. 2002. ``Humans and Consciousness." Perspectives on Science and Christian Faith 54, no. 1, 65-65.\\

Alexanian, Moorad. 2007. ``Debate about science and religion continues." Physics Today, 60, no. 2, 10, 12.\\

Alexanian, Moorad. 2007. ``Theistic Science: The Metaphysics of Science." Perspectives on Science and Christian Faith 59, no. 1, 85-86.\\

Alexanian, Moorad. 2006. ``Set Theoretic Analysis of the Whole of Reality." Perspectives on Science and Christian Faith 58, no. 3, 254-255.\\

Brooke, John. 1988. ``The God of Isaac Newton."  In Let Newton Be!, ed. R. Flood, J. Fauvel and R.J. Wilson, 169-83.  Oxford University Press.\\

Einstein, Albert. 2005. ``Physics, Philosophy, and Scientific Progress." Physics Today 58, no. 6, 46-48.\\

Jaynes, Edwin Thompson. 2003.  Probability Theory: The Logic of Science. Cambridge University Press.\\

Kahneman, Daniel, and Amos Tversky. 1973. ``On the Psychology of Prediction." Psychological Review 80:237-51.\\

O'Leary, Denyse. 2004.  By Design or by Chance?: The Growing Controversy on the Origins of Life in the Universe. Castle Quay.\\

Malebranche, Nicolas. 1997. ``The Search after Truth: With Elucidations of The Search after Truth (Cambridge Texts in the History of Philosophy)." Eds. Thomas M. Lennon and Paul J. Olscamp. Cambridge University Press.\\

Martin, William Oliver. 1957. The Order and Integration of Knowledge. The University of Michigan Press.\\

Nagel, Ernest and James R. Newman. 1958.  G\"{o}del's Proof.  New York University Press.\\

Pascal, Blaise. 1941. Pensees, and The Provincial Letters. Random House.\\

Thorson, Walter R.  2004. ``Naturalism and Design in Biology: Is Intelligent Dialogue Possible?" Perspectives on Science and Christian Faith 56, no. 1, 26-36.\\

Van Till, Howard J.  2004. ``Is the ID Movement Capable of Defeating Naturalism? A Response to Madden and Discher." Perspectives on Science and Christian Faith 56, no. 4, 292-295.\\

Wheeler, John A. and Kenneth W. Ford. 1998.  Geons, Black Holes, and Quantum Foam: A Life in Physics. W.W. Norton \& Company.\\

Wigner, Eugene P. 1999. ``On the Future of Physics." Fizikai Szemle 5: http://www.kfki.hu/fszemle/archivum/fsz9905/wigner.html\\

Wigner, Eugene P. 1960. ``The Unreasonable Effectiveness of Mathematics in the Natural Sciences." Communications on Pure and Applied Mathematics 13, no. 1, 1-14.\\

Wilder, T.E. 1991. ``At the Origins of English Rationalism." Contra Mundum, No.1, 2-16. Available at: http://www.contra-mundum.org/cm/cm01.pdf \\

\end{document}